\documentstyle[prl,aps,twocolumn,psfig]{revtex}

\catcode`\@=11

\def\maketitle2{\par % Uses \twocolumn[\@maketitle2].
\begingroup
\let\cite\@bylinecite
\def\thefootnote{\fnsymbol{footnote}}%
\twocolumn[\@maketitle2\vskip2pc]%
\thispagestyle{plain}\@thanks
\endgroup
\def\thefootnote{\arabic{footnote}}%
\setcounter{footnote}{0}%
\let\maketitle2\relax \let\@maketitle2\relax
\let\@thanks\relax \let\@authoraddress\relax \let\@title\relax
\let\@date\relax \let\thanks\relax \let\@abstract\relax 
\let\@pacs\relax}

\def\abstract#1{\gdef\@abstract{{\par % Store abstract text. 
\bgroup
\ifdim\prevdepth=-1000pt \prevdepth0pt\fi
\hsize\columnwidth
\dimen0=-\prevdepth \advance\dimen0 by17.5pt \nointerlineskip
\small\vrule width 0pt height\dimen0 \relax}{~~}#1\egroup}}

\def\pacs#1{\gdef\@pacs{{\par % Store PACS numbers as \@pacs.
\bgroup
\hsize\columnwidth \parindent0pt
\ifdim\prevdepth=-1000pt \prevdepth0pt\fi
\dimen0=-\prevdepth \advance\dimen0 by20pt\nointerlineskip
\egroup} PACS numbers:~#1}}

\def\@maketitle2{% Puts \@abstract and \@pacs in a {list}.
\@preprint
\@title
\ifdim\prevdepth=-1000pt \prevdepth0pt\fi
\@authoraddress
\@date
\begin{list}{}{\leftmargin=0.10753\textwidth \rightmargin=\leftmargin
\itemsep=1pc\partopsep=-1pc}
\item\@abstract
\item\@pacs
\end{list}
}

\catcode`\@=12

\begin{document}

\title{Deconstructing Decoherence}

\author{J.~R.~Anglin$^1$\thanks{anglin@lanl.gov}, 
J.~P.~Paz$^2$\thanks{paz@dfuba.df.uba.ar} and W.~H.~Zurek$^1$\thanks{whz@lanl.gov}}

\address{$^1$Theoretical~Astrophysics, T-6, Mail~Stop~B288, 
Los~Alamos~National~Laboratory, Los~Alamos, NM~87545}

\address{$^2$Departamento~de~Fisica, FCEN, UBA, Pabellon~1, 
Ciudad~Universitaria, 1428~Buenos~Aires, Argentina}

%\maketitle

%\vspace{1.0in}

\abstract{The study of environmentally induced superselection and of
the process of decoherence was originally motivated by the search for
the emergence of classical behavior out of the quantum substrate, in
the macroscopic limit\cite{Zurek8182}.  This limit, and other
simplifying assumptions, have allowed the derivation of several
simple results characterizing the onset of environmentally induced
superselection; but these results are increasingly often regarded as a
complete phenomenological characterization of decoherence in any regime.
This is not necessarily the case: The examples presented in this paper
counteract this impression by violating several of the simple ``rules
of thumb''.  This is relevant because decoherence is now beginning
to be tested experimentally\cite{Monroe,Brunetal}, and one may
anticipate that, in at least some of the proposed applications ({\it
e.g.,} quantum computers), only the basic principle of ``monitoring by
the environment''\cite{Zurek8182} will survive.  The phenomenology of decoherence may turn out to be significantly different.}


\maketitle2

\narrowtext

\section{Introduction}

According to deconstructionist philosophers, words refer only to other
words.  There is a certain amount of truth in the analogous suggestion
that papers in theoretical physics refer only to other papers (and quite
often, only to other papers in {\it theoretical} physics).
Consequently, a term like ``decoherence'' is in real danger of coming
to mean, to most physicists, only the processes that have been
most frequently studied in the literature.  Most of this literature has
heretofore dealt, naturally enough, with highly idealized models
amenable to exact solution.  Moreover, many of these models have been
particularly designed to
realize a macroscopic classical limit, in order to attain the original 
goal of understanding the quantum origins of classicality.  Such models have provided a relatively
small set of principles, which could easily be taken to govern
decoherence in general.  It is tempting, for example, to quote a simple
formula derived from a linear model\cite{CaldeiraLeggett,UnruhZurek} as
giving ``the'' decoherence timescale\cite{Zurek86}.
Emblematic of this problem is a well-known cartoon that appears in
introductory discussions of decoherence\cite{Zurek91}, depicting a
border crossing between the two realms of classical and quantum
physics.  While this is a provocative metaphor, it may prompt the inaccurate
impression that there is exactly one well-defined way of crossing from
one realm to the other.

In this paper we will effectively argue that many perceived universalities
in the phenomenology of decoherence are artifacts of studying toy
models, and that the single neat border checkpoint should be replaced
as an image for decoherence by the picture of a wide and ambiguous
No Man's Land, filled with pits and mines, which may be crossed on
a great variety of more or less tortuous routes.  Once one has indeed
crossed this region, and travelled some distance away from it, the going
becomes easier:  we are not casting doubt on the ability of the very
strong decoherence acting on macroscopic objects to enforce effective
classicality.  But in the near future precise experiments (for example,
\cite{Brunetal,Monroe,Schmied,Ekstrom,CTetal,Haroche,CiracZoller,Turchette})
will explore regimes in which decoherence should be measurable, but
not so strong as simply to enforce classicality.  Experiment is thus
beginning to probe the quantum/classical No Man's Land itself, advancing
daring patrols along an impressively broad front.  In comparing the
results of these experiments with theoretical predictions, it will be
important not to assume that the simple cases examined so far should
be taken as representative of decoherence in general.  By presenting
a number of theoretically tractable examples in which various elements
of phenomenological lore can be seen to fail explicitly, we make the point
that each experimental scenario will have to be examined theoretically
on its own merits, and from first principles.

{}From the bulk of previous theoretical studies of decoherence, one
might be tempted to deduce three significant principles concerning the
rate of decoherence: one can define a simple decoherence time scale
which is valid at least for linear systems at high temperature; the
rate of decoherence of classically impossible ``Schr\"odinger's Cat''
states is always set by the fastest time scales present; and the rate of
decoherence increases with the square of the distance between the two
branches of such Cat states.  These elements of the standard lore are
indeed borne out in the results of the first decoherence experiment
at hand\cite{Brunetal}; but there is no guarantee that they will
always hold.  We therefore show why in the most general mesoscopic
regime one may need to go back to the basic idea that the environment
``monitors'' an open quantum system\cite{Zurek8182}, and from there
derive phenomenology afresh for every model.  We will consider the three
putative principles in successive sections, presenting in each section
an explicit example in which the property determined for simple models
previously studied no longer holds.  A final section will then discuss
our results collectively, and suggest some implications of them for the
interpretation of experiments currently proposed or in progress.

\section{Decoherence timescale in linear Brownian motion}

Many studies of decoherence have involved completely linear models, in
which a single Brownian particle is placed in a quadratic potential,
and coupled linearly to a heat bath composed of (often, uncountably
many) harmonic oscillators.  It can in fact be argued\cite{AnglinZurek}
that environments with non-linear internal dynamics can often be
closely approximated, as far as their effects on the observed system
are concerned, by such an independent oscillator model.  Although there
are certainly cases in which it is not realistic, the independent
oscillator model is therefore not entirely a toy, and represents a
simplicity that is actually realized in nature.  And simple as it is,
even it is not really simple as special cases and convenient approximations
often make it appear.

The canonical example of decoherence is the evolution of a Brownian
harmonic oscillator from an initial state which is a superposition of
two coherent states localized at distinct positions in space.  This
initially pure state, assumed to be uncorrelated with the initial
thermal state of an independent oscillator environment, has been found
to evolve rapidly into an incoherent mixture of the two coherent
states.  Simple formulas are often applied to quantify ``rapidly''.
Here, however, we will present an easy derivation of the short-time
behaviour of the Wigner function for an Ohmic Brownian oscillator, and
show that there is in general no natural way to identify a single time
scale for decoherence, even in the high temperature limit.  Our more
explicit results are in agreement with the physical conclusions reached
on the basis of numerical evidence in Reference \cite{PHZ}.

For our completely linear model, we take the Hamiltonian
\begin{eqnarray}\label{linHam}
H &=& {1\over2M}P^2 + {M\Omega^2\over2}Q^2\nonumber\\
& &\; +{1\over2}\int_0^\infty\!d\omega\,
	[(p_\omega +gf_\omega Q)^2+\omega^2 q_\omega^2]\;,
\end{eqnarray}
where $P$ and $Q$ are the Brownian particle's canonical variables, and
$M$ and $\Omega$ are its mass and natural frequency; $p_\omega$ and
$q_\omega$ are the canonical variables for the bath oscillator with
frequency $\omega$; $g$ is an overall coupling strength which may be used
to define the dissipation rate
\begin{equation}\label{gamma0}
\gamma \equiv {\pi g^2\over4M}\;;
\end{equation}
and $f_\omega$ describes the relative coupling strength of the various
environmental modes.  The square of this strength will play
the role of a spectral density.

The initial Wigner function $W(Q,P;0)$ of the Brownian oscillator will
be that for an equal amplitude superposition of two coherent states,
whose wave functions are Gaussians displaced equal and opposite amounts
$\pm a$ from the origin.  This Wigner function contains two terms,
then: one consisting of a sum of two Gaussians, representing the
incoherent mixture of the two states; and one which is oscillatory, and
represents their quantum interference:
\begin{eqnarray}\label{Wiginit}
W(Q,P;0) &=& W_{mix} + W_{int}\nonumber\\
W_{mix}(Q,P;0) &=& {(1-e^{-M\Omega a^2/\hbar})^{-1}\over\pi\hbar}
	\cosh\Bigl(2{M\Omega\over\hbar}aQ\Bigr)\nonumber\\
&&\times \exp\Bigl[-{1\over\hbar\Omega}
	\Bigl({P^2\over M} + M\Omega^2 (Q^2 + a^2)\Bigr)\Bigr]
	\nonumber\\ 
W_{int}(Q,P;0) &=& {(1-e^{-M\Omega a^2/\hbar})^{-1}\over\pi\hbar}
	\cos\Bigl(2{aP\over\hbar}\Bigr)\nonumber\\
&&\times \exp\Bigl[-{1\over\hbar\Omega}
	\Bigl({P^2\over M} + M\Omega^2 Q^2\Bigr)\Bigr]
\;.
\end{eqnarray}
Decoherence in this model appears as a rapid decay in magnitude of
$W_{int}(Q,P;t)$, by means of an exponential prefactor $e^{-D(t)}$.

The initial Wigner function for the complete system of Brownian
oscillator plus bath is assumed to be a direct product
\begin{eqnarray}\label{WW}
\lefteqn{{\cal W}(Q,P;\{q_\omega,p_\omega\};0) = W(Q,P;0)\times W_e[q_\omega,p_\omega]}\nonumber\\ 
&& W_e[q_\omega,p_\omega] = \prod_\omega 
	{\tanh(\hbar\beta\omega /2)\over\pi\hbar}\nonumber\\
&& \qquad\qquad\times \exp\Bigl[-{1\over\hbar\omega}
	(p^2_\omega + \omega^2 q^2_\omega)
	\tanh{\hbar\beta\omega\over2}\Bigr]\;,
\end{eqnarray}
where $\beta = (k_B T)^{-1}$ is the inverse temperature of the
environment. 

It can be shown quite easily that the Wigner function for a
totally linear system evolves under the same Liouville equation as the
classical ensemble density for the same model.  Consequently, we can
evolve the Wigner function by simply propagating it along the classical
trajectories in phase space.  The reduced Wigner function for the
Brownian particle alone, with the environment integrated out, is
therefore
\eject
\begin{eqnarray}\label{Wigt}
\lefteqn{W(Q_F,P_F;t_F)}\nonumber\\
 &=& \int\!dQ_IdP_I{\cal D}q_{\omega I}{\cal D}p_{\omega I}\,
	\delta\bigl(Q_F-Q_0(t_F)\bigr)
	\delta\bigl(P_F-P_0(t_F)\bigr)\nonumber\\
& &\qquad\times W(Q_I,P_I;0) W_e[q_{\omega I},p_{\omega I}]\\
&=& \int {\cal D}q_{\omega I} {\cal D}p_{\omega I}\,
\Bigg\vert{\partial (Q_I,P_I)\over\partial (Q_F,P_F)}\Bigg\vert\,
W_e[q_{\omega I},p_{\omega I}]\,W\bigl(Q_I,P_I;0)\;,\nonumber
\end{eqnarray}
where $Q_0(t)$ and $P_0(t)$ are given by Hamilton's equations, for the
Hamiltonian (\ref{linHam}).  We have simplified presentation in
(\ref{Wigt}) at the expense of precise notation:  in the first line,
$Q_I$ and $P_I$ are dummy variables, and we implicitly assume the
initial boundary conditions $Q_0(0) = Q_I, P_0(0)=P_I$; but in the
second line, we intend instead the final boundary conditions
$Q_0(t)=Q_F, P_0(t) = P_F$, and we use $Q_I,P_I$ as shorthand for the
resulting $Q_0(0),P_0(0)$.  In the remainder of this discussion, we
will continue the usage of the second line, according to which it
should be noted that $Q_I$ and $P_I$ are in fact functions of the final
time $t_F$, and linear functions of $Q_F$, $P_F$, and initial
environmental variables $\{q_{\omega I}, p_{\omega I}\}$.

We are interested in decoherence that occurs on time scales much shorter
than the Brownian particle's dynamical time scale $\Omega^{-1}$,
and when the environment is very weakly coupled to the system.  We will
therefore solve the equations of motion for $Q_0$ and $P_0$ perturbatively
to first order in $\Omega t$ and at most first order in $g$, to obtain
\begin{eqnarray}\label{firstorder}
Q_I(t) &\doteq& Q_F - {P_F\over M}t \nonumber\\
P_I(t) &\doteq& P_F + M\Omega^2 Q_F t + \int_0^t\!dt'\,F_1(t')\;,
\end{eqnarray}
where $F_1(t)$ is the force exerted by the environment, to first order
in $g$.  Since this force will be a linear function of the $q_{\omega
I}$ and $p_{\omega I}$, and since to form the reduced Wigner function
$W(Q_F,P_F;t_F)$ we will be integrating over these variables with the
Gaussian weight $W_e$, Eqn.  (\ref{firstorder}) is effectively a
Langevin equation with a Gaussian stochastic force.  Note also that
Equation (\ref{firstorder}) implies that the Jacobian in Equation
(\ref{Wigt}) is simply $1$, to first order in $\Omega t$.

There are some subtle points to be considered before writing down the
expression for $F_1(t)$.  One might be tempted simply to write $F_1(t) 
\doteq F_1(0) =
g\int\!d\omega\,f_\omega p_{\omega I}$; but
this would be forgetting the fact that $F_1(t)$ can contain some
frequencies much higher than $\Omega$, so that some components of the
stochastic force will oscillate significantly even over the short time
interval in which we can expect to see decoherence.  We therefore
write the more accurate expression
\begin{equation}\label{Ohmacc}
F_1(t) = g\int_0^\infty\!d\omega\,f_\omega\,[p_{\omega I} 
	\cos\omega t + \omega q_{\omega I}\sin\omega t]\;.
\end{equation}

Actually, neglecting higher order terms in $g$ will be inaccurate, even
for very early times, if the high-frequency end of the environmental
spectrum is too strong.  As one finds by fully solving such
``supra-ohmic'' models, higher order terms in $g$ can appear multiplied
by large frequencies, and thus be significant.  In such cases,
backreaction can be so swift that a counterterm to the ``bare'' force
$F_1(t)$ is generated rapidly enough to affect (and typically suppress)
decoherence.  One can understand this phenomenon roughly as the rapid
onset of adiabatic dragging of the high frequency bath degrees of
freedom; it is discussed in detail, in Reference \cite{Soqbm}.

These subtleties of backreaction turn out to be insignificant in the
much-studied Ohmic case, where (for the coupling scheme we are using)
$f_\omega$ is constant up to some high UV cut-off scale.  We will
therefore assume the Ohmic case, choosing for definiteness the
Lorentzian cut-off scheme
\begin{equation}\label{Ohmf}
f_\omega = {\Gamma\over\sqrt{\omega^2+\Gamma^2}}
\end{equation}
with $\Gamma >> \Omega$, 
and accept Equation (\ref{Ohmacc}) as valid.  Working to first order in
$\Omega t$, we find that the Brownian particle gains negligible
energy from the environment at these very early times: 
\begin{equation}\label{en}
{P_I^2\over M} + M\Omega^2Q_I^2 \doteq {P_F^2\over M} + M\Omega^2Q_F^2\;,
\end{equation}
when we neglect $g$ completely because we assume that
$P_I\int\!dt'\,F(t')/M$ is negligible for the
$|P_I|\sim\sqrt{M\hbar\Omega}$ that are significant in $W(Q_I,P_I;0)$.
Even though the environmental force is too small to affect the energy
of the Brownian particle at these early times, however, $a>>\sqrt{\hbar
/M\Omega}$ will allow the change in $aP$ to be significant:
\begin{eqnarray}\label{aP}
aP_I &\doteq& aP_F\\ 
& &\;+ag\int_0^\infty\!{d\omega\over\omega}f_\omega\,
	[p_{\omega I}\sin\omega t 
	+ \omega q_{\omega I}(1-\cos\omega t)]\;.\nonumber
\end{eqnarray}
Performing the Gaussian integrals in Equation (\ref{Wigt}) using
(\ref{en}) and (\ref{aP}), we find that $W_{mix}(Q,P;t)$ is negligibly
changed from $W_{mix}(Q,P;0)$, but that $W_{int}(Q,P;0)$ has evolved
into
\begin{equation}\label{Wmixt}
W_{mix}(Q,P;t) \doteq e^{-D(t)} W_{mix}(Q,P;0)\;,
\end{equation}
where the decoherence factor $D(t)$ is given by
\begin{equation}\label{Dt}
D(t) = {8M\gamma a^2\over\pi\hbar}\int_0^\infty\!{d\omega\over\omega}\, f_\omega^2\coth{\hbar\beta\omega\over2}\,(1-\cos\omega t)\;.
\end{equation}

In the zero temperature limit, Equation (\ref{Dt}) agrees with Eqns.
(36)--(37) of Reference \cite{decoreco}, which present a weak coupling,
early time approximation to an exact solution once it has been
obtained.  In the high temperature limit, we can explicitly evaluate
$D(t)$ as
\begin{equation}\label{hightemp}
D(t) \stackrel{\scriptscriptstyle T\to\infty}{\longrightarrow}
	{8\gamma k_B T a^2\over\hbar^2}
	\Bigl(t-{1-e^{-\Gamma t}\over\Gamma}\Bigr) \;.
\end{equation}
For times much less than $\Omega^{-1}$ but still much greater
than $\Gamma^{-1}$, Equation (\ref{hightemp})
agrees with previous results that at high temperatures $D(t) \propto t$.
This linear behaviour of $D(t)$ allows one to specify a single decoherence
time scale
\begin{equation}\label{tdec}
\tau_{dec} = {\hbar^2\over 8M\gamma a^2 k_B T}\;.
\end{equation}
Even when the high temperature limit $k_BT>>\hbar\Gamma$ is valid,
however, this formula is not really universal.  For sufficiently high
$T$ or $a^2$, decoherence will already have occurred ($e^{-D(t)}<<1$) at
times smaller than or on the order of $\Gamma^{-1}$.  We will then have
to write 
\begin{equation}\label{tdec'}
D(t) \simeq {4M\gamma k_BT a^2 \over\hbar^2}\Gamma t^2\;,
\end{equation}
from which one must deduce the much longer timescale
\begin{equation}
\tau'_{dec} = {\hbar\over 2a\sqrt{M\gamma\Gamma k_BT}}\;.
\end{equation}

For lower temperatures, or non-Ohmic environments, $D(t)$ will
generally not be linear, and the time at which $e^{-D(t)} << 1$ will be
a complicated function of temperature and $a^2$.  The existence of a
single simple formula for ``the'' decoherence time scale is a special
property of the Ohmic independent oscillator model at high, but not
ultra-high, temperatures.

\section{Initial state preparation}

Simple or not, all the decoherence timescales which might be identified
in models like that of Section II have the common feature of being very
short.  Warnings have long been made, however, that that the rapidity
of this initial burst of decoherence might be spurious, in that it
might be a special consequence of an initial state in which the system
and environment are negligibly entangled.  Since it is the high
frequency modes of the environment that are responsible for rapid
decoherence, the neglect of initial entanglement is particularly
dubious: these fast modes are precisely the ones which will tend to be
adiabatically dragged along with the system, if the system is put into
a ``Schr\"odinger's Cat'' state by a physical process instead of by
theoretical {\it fiat}.  Despite warnings about this issue in prose,
however, there has so far been no actual calculation to really lay this
ghost to rest.

In this section we examine a model which is essentially the same as
those of Section II or Reference \cite{decoreco}.  Instead of following
the evolution of an initial superposition of displaced Gaussian states,
however, we will take the ground state of the complete system as our
initial state, and apply an external force which drives the Brownian
oscillator into a superposition of displaced Gaussians over a finite
period of time.  We find that decoherence occurs in this scenario, but
that it is no longer characterized by the short UV time scale.  The
strong initial burst of decoherence, which has been ubiquitous but
suspect in previous studies, is indeed suppressed.

We again take the Hamiltonian
\begin{eqnarray}\label{linHam2}
H_0 &=& {P^2\over2M} + {M\Omega^2\over2}Q^2\nonumber\\ 
& &\; +{1\over2}\int_0^\infty\!d\omega\,
	[(p_\omega +gf_\omega Q)^2+\omega^2 q_\omega^2] \;,
\end{eqnarray}
just as in Equation (\ref{linHam}) above.  We also retain the Ohmic specification
for $f_\omega$ given by Equation (\ref{Ohmf}).  We do make an important
change in our system, however, even though it does not show up in
$H_0$: we endow our Brownian oscillator with a two-state internal
degree of freedom, such as a spin.  The Hamiltonian as written so far
does not distinguish between the oscillator's two internal states; but
we now add to it an external force which does distinguish them, and
which will thereby be able to create a Schr\"odinger's Cat state from
the ground state:
\begin{equation}\label{extforce}
H_{\alpha} = H_0 + a\alpha(t) \hat\sigma {P}\;.
\end{equation} 
Here $a$ is again a distance scale, $\alpha(t)$ is a time-dependent
c-number having dimensions of frequency, with $\alpha(0) = 0$, and the
Pauli spin matrix $\hat\sigma$ acts in the internal space.  We will
then take our initial state to be
\begin{equation}\label{psiinit}
|\Psi_i\rangle = {1\over\sqrt2}|\phi_0\rangle \bigl(|+\rangle + |-\rangle\bigr)\;,
\end{equation}
where $|\phi_0\rangle$ is the ground state of $H_0$, and $\hat\sigma
|\pm\rangle = \sigma |\pm\rangle$ for $\sigma = \pm 1$.

Since the internal state of the oscillator does not evolve in this
model, the two different realizations of $\sigma$ which are present in
the initial state merely label two branches of the total quantum state
at any time.  For non-zero $\alpha(t)$, the spatial wave functions
associated with these two branches will over time become quite
different.  Choosing $\alpha(t) = 2\delta(t)$, for example, will
reproduce the initial Schr\"odinger's Cat state of Reference
\cite{decoreco} (which is very similar to that of Section II above).
In what follows here we will consider the case where $\alpha(t)$ is not
a delta function.

As explained in Reference \cite{decoreco}, $H_0$ can be diagonalized
by defining new operators $A_\omega, \pi^A_\omega$:
\begin{equation}\label{Hdiag}
{H}_0 = {1\over2}\int_0^\infty\!d\omega\,\bigl[(\pi^A_\omega)^2 
	+ \omega^2 A_\omega^2\bigr]\;,
\end{equation}
where 
\begin{eqnarray}\label{Pom}
{P} &=& \int_0^\infty\!d\omega\,p(\omega) {A}_\omega\\
p(\omega) &=& {g\omega^2\Gamma\over\sqrt{\pi
	[\omega^2 + \bar\Gamma^2][\omega^2 
		- (\bar\Omega+i\bar\gamma)^2]
	[\omega^2 - (\bar\Omega -i\bar\gamma)^2]}}\;.\nonumber
\end{eqnarray}
The barred quantities $\bar\Gamma$, $\bar\Omega$, and $\bar\gamma$ are
renormalized versions of the bare parameters.  The bare parameters may
be expressed simply in terms of the renormalized ones (the inverse
relation being a complicated cubic formula)\cite{decoreco}, but we will
assume that $\Gamma >> \Omega >> \gamma$, and in this case the
differences between the barred and unbarred quantities are negligible.
${Q}$, ${q}_\omega$, and ${p}_\omega$ may also be expressed
in terms of the new operators, but we will only be needing Eqn.
(\ref{Pom}).

Since the wave functional for the ground state $|\phi_0\rangle$ is the
familiar harmonic oscillator Gaussian, it is easy to work out the wave functional for the
state at time $t$ in the $\pi^A_\omega$ representation:
\begin{eqnarray}\label{psipi}
\lefteqn{\Psi[\pi^A_\omega,\sigma;t]}\nonumber\\
&=& \langle\sigma|\langle\pi_\omega^A|
	{\cal T}e^{-{i\over\hbar}\int_0^t\!dt'\,{H}_\alpha(t')}
	|\phi_0\rangle |\sigma\rangle \nonumber\\
&=& Z(t) \exp\Bigl[-{1\over2\hbar}\int_0^\infty\!{d\omega\over\omega}\,
	\Bigl([\pi^A_\omega\nonumber\\
& &\qquad + \sigma p(\omega)
		\int_0^t\!dt'\,\alpha(t')\cos\omega(t-t')]^2\nonumber\\
&&\qquad +2i\sigma p(\omega) \pi^A_\omega 
		\int_0^t\!dt'\,\alpha(t')\sin\omega(t-t')\Bigr)\Bigr]\;.
\end{eqnarray}
${\cal T}$ denotes time ordering, and $Z(t)$ is a normalization constant
into which we have absorbed an irrelevant time-dependent phase.  We can
then obtain the reduced density matrix for the Brownian particle, in the
$Q$ representation, merely by performing some Gaussian integrals:
\begin{eqnarray}\label{RDM}
\lefteqn{\rho(Q,Q',\sigma,\sigma';t)}\nonumber\\
&=& \int\!d\xi\int\!{\cal D}\pi^A\,
	\exp\Bigl[{i\over\hbar}\xi\int_0^\infty\!d\omega 
	{p(\omega)\over M\omega^2} \pi^A_\omega\Bigr]\nonumber\\
& & \qquad\qquad \times
	\Psi[\pi^A_\omega-p(\omega)Q,\sigma;t]
	\Psi^*[\pi^A_\omega-p(\omega)Q',\sigma';t]\nonumber\\
&=& N \exp\Bigl[-{M\Omega_2\over4\hbar}\Bigl(
	{\Omega_1\over\Omega_2}
	[Q-Q' \nonumber\\
& & \qquad\qquad\qquad\qquad 
	- (\sigma-\sigma')\int_0^t\!dt'\alpha(t')y(t-t')]^2
	\nonumber\\
& & \qquad\qquad
	+[Q+Q' - (\sigma+\sigma')\int_0^t\!dt'\alpha(t')r(t-t')]^2
	\nonumber\\
& &\qquad\qquad -2i(\sigma+\sigma')(Q-Q')\int_0^t\!dt'\alpha(t')s(t-t')
	\nonumber\\
& &\qquad\qquad -2i(\sigma-\sigma')(Q+Q')
	\int_0^t\!dt'\alpha(t')z(t-t')\Bigr]
	\nonumber\\
& &\times \exp\Bigl[-{(\sigma-\sigma')^2\over4}D_\alpha(t)\Bigr]\;.
\end{eqnarray}

Several new functions and quantities have been introduced in Eqn.
(\ref{RDM}). $N$ is simply a normalization constant.  There are two new
frequencies
\begin{eqnarray}\label{Om12}
\Omega_1 &\equiv& {1\over M}\int_0^\infty\!d\omega\,
	{[p(\omega)]^2\over\omega}\nonumber\\
\Omega_2 &\equiv& M\left[\int_0^\infty\!d\omega\,
	{[p(\omega)]^2\over\omega^3}\right]^{-1}\;.
\end{eqnarray}
Using these we also define four dimensionless functions
\begin{eqnarray}\label{functions}
r(t) &\equiv& {1\over M}\int_0^\infty\!d\omega\,
	{[p(\omega)]^2\over\omega^2}\cos\omega t\nonumber\\
s(t) &\equiv& {1\over M\Omega_2}\int_0^\infty\!d\omega\,
	{[p(\omega)]^2\over\omega}\sin\omega t\nonumber\\
y(t) &\equiv& {1\over M\Omega_1}\int_0^\infty\!d\omega\,
	{[p(\omega)]^2\over\omega}\cos\omega t\nonumber\\
z(t) &\equiv& {1\over M}\int_0^\infty\!d\omega\,
	{[p(\omega)]^2\over\omega^2}\sin\omega t\;.
\end{eqnarray}
Note that $r(0)=y(0)=1$, and $s(0)=z(0)=0$.  These functions may all be
evaluated explicitly by contour integration.  One finds that $r(t)$ and
$s(t)$ are (for $\Gamma>>\Omega>>\gamma$) very close to $e^{-\bar\gamma
t}\cos\bar\Omega t$ and $e^{-\bar\gamma t}\sin\bar\Omega t$,
respectively, while $y(t)$ and $z(t)$ are similar, but also include
some exponential-integral terms (at first order in $(\gamma/\Omega)$).
We can therefore see that (\ref{RDM}) prescribes evolution of Gaussian
peaks along classical trajectories, for the ``diagonal'' terms with $\sigma=\sigma'$.  The interference terms, with $\sigma=-\sigma'$, 
evolve slightly differently, but are also suppressed by the decoherence
prefactor $e^{-D_\alpha(t)}$.

This prefactor is given by
\begin{eqnarray}\label{Dalphat}
D_\alpha(t) &=& M\Omega_1\Bigl[\int_0^t\!dt'\!\int_0^t\!dt''\,\alpha(t')\alpha(t'') 
	y(t'-t'')\nonumber\\
& &\qquad\qquad -\Bigl(\int_0^t\!dt'\,
		\alpha(t')y(t-t')\Bigr)^2\Bigr]\nonumber\\
& &\qquad\qquad - M\Omega_2\left(\int_0^t\!dt'\,
	\alpha(t')z(t-t')\right)^2\;.
\end{eqnarray}
In the case where $\alpha(t) = 2\delta(t)$, decoherence is rapid because
the function $1-y^2(t)$ grows on the cut-off timescale $\Gamma^{-1}$.
This occurs because, as one can see by inserting (\ref{Pom}) in
(\ref{functions}), $\Omega_1 y(t)$ diverges logarithmically when
$\Gamma\to\infty$ and $t\to0$.  Hence $\Omega_1 y(t)$ drops precipitously
within a few cut-off times of $t=0$.  But the convolutions appearing
in (\ref{Dalphat}) clearly cannot vary more rapidly than $\alpha(t)$
itself.  If one chooses $\alpha(t) = \sin\Lambda t$ for some $\Lambda
<< \Gamma$, for example, the logarithmic divergence in $\Omega_1 y(t)$
for $t\to0$ will be regulated by the smearing with $\alpha(t)$, and
nothing in $D_\alpha(t)$ will evolve on a time scale set by $\Gamma$.
We can therefore see that, if a Schr\"odinger's Cat state is created by
some physical process (as in Refs.~\cite{Monroe} and \cite{Brunetal}), rather than by
theorist's {\it fiat}, the rate of decoherence will no longer be set by
the cut-off scale, but instead by some combination of the timescales of
$\alpha(t)$, $\Omega$, and $\gamma$.  In general, an upper bound on the
decoherence time scale is set by the time scale on which a Schr\"odinger's
Cat state is actually constructed in the laboratory.

\section{Saturation of decoherence at long range}

In both of the examples we have studied to this point, the decoherence
exponent $D(t)$ scales quadratically with the separation scale $a$.
In this section, we consider two cases in which a single particle which
interacts non-linearly (quasi-locally) with a linear environment, and
the rate of decoherence of two localized states of the particle turns
out not to increase indefinitely with the distance between the two
particle positions.  Instead the decoherence rate reaches a plateau at
some distance, which is set by the range of the interaction between the
particle and the environment.  

This point has been argued persuasively by Gallis and Fleming
\cite{GallisFleming} and by Gallis \cite{Gallis1,Gallis2}, in several
insightful papers.  At the level of general principle, the calculations
we present in this Section supplement and support their results.  We
are able to proceed somewhat further, however, both in solving a simple
model exactly, and in deriving results from first principles without
phenomenological assumptions.  At a more detailed level, our results
differ from those of Gallis and Fleming, in that we identify cases
where the lengthscale at which decoherence saturates is set not by an
environmental correlation length, but by an interaction range, or by
the time over which the interaction occurs.

The first of our cases is an idealized model which can be solved
exactly (in the sense that the evolution of the quantum state is
determined by a non-linear first order {\it ordinary} differential
equation, which can itself be solved analytically in some non-trivial
cases).  The second is a more realistic model, in which the environment
is a quantum field, but we will only be able to describe certain
features of the influence functional that are clearly relevant to
decoherence.

\subsection{The `mattress model'}

We consider a non-relativistic quantum particle in one dimension, which
is free except for its interaction with an environment.  This
environment resembles an expensive (but one-dimensional) mattress: it
consists of a series of independent `pocketed coil' spring-systems,
sited at equal intervals along a line, each interacting with the
particle only when it is sufficiently near to them.  The Lagrangian for
this system is
\begin{eqnarray}\label{Lmat}
L_{mat} &=& {M\over2}\dot{x}^2 + 
	{1\over2}\sum_{n=-N}^N\int_0^\infty\!d\omega\,I(\omega)
	\Bigl(\dot{q}_{n,\omega}^2\nonumber\\
& & \qquad\qquad\qquad - \omega^2[q_{n,\omega}-
	{g\over\omega}f(x-nd)]^2\Bigr)\;,
\end{eqnarray}
where $M$ is the particle mass, $x$ is its position in space, $n$
labels the $2N+1$ sites of the `pocketed coils', and $d$ is the
distance between these sites.  Each pocketed coil consists of a number
of linear springs whose displacements are $q_{n,\omega}$, having
natural frequencies $\omega$, distributed according to the spectral
density $I(\omega)$.  The springs are connected to the particle with a
coupling strength $g$, modulated by the spatial profile $f(x)$.  By our
prescription that the interaction be `quasi-local', we mean that we
will assume that $f(x)$ vanishes for $|x|\to\infty$.

The evolution of the reduced density matrix of the Brownian particle is
expressed in path integral language as
\begin{eqnarray}\label{RDMevol}
\rho(x_f,x'_f;t) &=& \int\!dx_idx'_i\,\rho(x_i,x'_i;0)\nonumber\\
& &\;\times
	\int\limits_{x_i}^{x_f}\!{\cal D}x\!
	\int\limits_{x'_i}^{x'_f}\!{\cal D}x'\,
	e^{{i\over\hbar}(S[x]-S[x'])} F[x,x']\;,
\end{eqnarray}
where $F[x,x']$ is the {\it influence functional}.  Since the
environment in this model is merely a collection of harmonic
oscillators, it is easy to compute $F[x,x']$.
If we take $I(\omega)$ to be a constant up to some irrelevantly large
cut-off frequency $\Gamma_m$, and assume that the environment is
initially in a high temperature ($k_B T>> \hbar\Gamma_m$) thermal state,
uncorrelated with the particle, we obtain for the influence functional 
the well known form
\begin{eqnarray}\label{wkf}
\lefteqn{F[x,x'] =}\nonumber\\
&& \exp\Bigl[-{g^2\over2\hbar}\int_0^t\!dt'\sum_{n=-N}^N
	\Bigl({k_BT\over\hbar}[f(x-nd)-f(x'-nd)]^2\nonumber\\
& & \qquad + i \delta(t')[f^2(x-nd) -f^2(x'-nd)]\nonumber\\
& & \qquad +{i\over2}[f(x-nd)-f(x'-nd)]\nonumber\\
&& \qquad\qquad\times
	[\dot{x}f'(x-nd) + \dot{x}'f'(x'-nd)]\Bigr)\Bigr]\;.
\end{eqnarray}

If we further take the infinite continuum limit $N\to\infty, d\to 0$,
and also let $g\to 0$ but keep constant $\mu \equiv {g^2\over4d}$,
we obtain the very simple case in which the evolution of the reduced
density matrix of the particle is given by the path integral
\begin{eqnarray}\label{PImat}
\lefteqn{\rho(x_f,x'_f;t) =}\nonumber\\
&&\; \int\!dx_idx'_i\,\rho(x_i,x'_i;0)\nonumber\\
&&\;\times\int\!{\cal D}\Delta{\cal D}\Sigma\, 
	\exp\Bigl[{i\over\hbar}\int_0^t\!dt'\,
	[M\dot{\Delta}\dot{\Sigma} 
	- 2\mu\dot{\Sigma} U'(\Delta)\nonumber\\
&&\qquad\qquad\qquad
	+ 4i{\mu k_B T\over\hbar} U(\Delta)]\Bigr]\;,
\end{eqnarray}
with the boundary conditions $\Delta(0)=x_i-x_i'$, $\Delta(t)=x_f-x_f'$,
$\Sigma(0) = (x_i+x_i')/2$ and $\Sigma(t) = (x_f+x_f')/2$, and where 
\begin{equation}\label{Fmat}
U(\Delta) \equiv \int_{-\infty}^\infty\!dy\, f(y)[f(y) - 
f(y - \Delta)]\;.
\end{equation}
As an example to indicate the implications of (\ref{Fmat}), note that a
Gaussian $f(y) \propto \exp[-a y^2]$ implies $U(\Delta) \propto
(1-\exp[-a\Delta^2/2])$ By analogy with the much-studied linear cases,
$U(\Delta)$ may be said to represent environmental noise acting on the
particle.  The fact that its derivative appears in Equation
(\ref{PImat}) as a dissipative term may be considered a
fluctuation-dissipation relation.  In the limit where $\mu\to0$ but
$T\to\infty$ so that $\mu T$ remains finite, we obtain the
dissipationless model of Gallis and Fleming\cite{GallisFleming}.  One
can therefore consider the present Section to be an extension of their
model into a regime in which a fluctuation-dissipation relation
exists.
 
Markovian dynamics, and the translation invariance that obtains in the
continuum limit, have conspired to make the exponent in
Equation (\ref{PImat}) linear in $\Sigma(t')$.  Consequently, the path
integral may be performed trivially, and we obtain the propagator
equation 
\begin{eqnarray}\label{propmat} 
\rho(x_f,x'_f;t) &=&
N(t)\int\!dx_idx'_i\,\Bigl(\rho(x_i,x'_i;0)\nonumber\\
& & \quad\times \exp\Bigl[{i\over\hbar} 
	{K\over2}(x_f+x'_f - x_i -x'_i)\Bigr]\nonumber\\
& & \quad\times \exp\Bigl[-4{\mu k_BT\over\hbar^2}
	\int_0^t\!dt'\,U(\Delta_0)\Bigr]\Bigr)\;, 
\end{eqnarray} 
where $N(t)$ is a normalization constant that is a relic of the path
integral measure.  $K=K(x_f-x'_f,x_i-x'_i,t)$ and $\Delta_0(t')$ are
defined by the promised first order ODE:
\begin{equation}\label{EofMmat} 
M\dot{\Delta}_0(t') - 2\mu
U'\bigl(\Delta_0(t')\bigr) = K\;, 
\end{equation} 
with $K=K(\Delta_f,\Delta_i,t)$ fixed by the two boundary
conditions $\Delta_0(t) = x_f-x'_f$ and $\Delta_0(0) = x_i - x'_i$.

We pause here to summarize our results so far.  We have considered a
model in which, in effect, every point in one dimensional space holds
an independent oscillator heat bath, which provides Ohmic dissipation
and white noise to a free particle, as long as it is within range.
This model thus represents a conveniently ideal limit of any scenario
in which a particle interacts locally with its environment, and
information transport within this environment is negligible.  As with
totally linear models, the path integral for this open quantum system
can be performed analytically; but this model contains non-linear
dynamics, in the coupling profile $f(x)$.  We now proceed to
investigate some consequences of this non-linearity.

{}From the assumption that $f(x)$ vanishes for large $|x|$, we can
easily derive certain properties of the important overlap function
$U(\Delta)$.  By examining Equation (\ref{Fmat}) in Fourier space, we
can see that $U(\Delta) > 0$, except at $\Delta=0$.  $U$ thus clearly
drives decoherence of superpositions of quantum states that are
localized at different locations.  Furthermore, one can easily show
that $U(0) = U'(0) =0$, and that $U''(0) > 0$.  For small $\Delta$,
then, $U$ looks like a parabola.  If we were to take $U$ to be a
parabola exactly, however, we would obtain merely the high temperature
limit of the free-particle Caldeira-Leggett model\cite{CaldeiraLeggett}.\footnote{Since the
Caldeira-Legget model is dynamically classical, it is not surprising
that the dynamics of the classical mattress model for any $f(x)$ is
also only sensitive to $U''(0)$, and not to $U(\Delta)$ as a whole.}
But we can also see from Eqn.  (\ref{Fmat}) that for large $\Delta$,
$F(\Delta)$ approaches the positive constant $\int\! dy\, f^2(y)$ ---
which may be set equal to 1 by rescaling $\mu$.  This saturating
behaviour of the decoherence term is arguably a generic effect of
locally coupled environments:  states of the environment that are
deformed differently by interaction with the particle at different
locations are just as orthogonal if these two locations are barely out
of interaction range with each other, as if they were infinitely far
apart.  A miss is as good as a mile.

By establishing the saturation of decoherence with increasing distance,
we have attained the real point of this subsection.  As an interesting
appendix, though, we point out that we can actually proceed further in
solving the mattress model, by constructing the $(k,\Delta)$
representation of the density matrix --- the ``Rengiw function''
$R(k,\Delta)$.
\begin{equation}\label{rengiw}
\rho(\Sigma+{\Delta\over2},\Sigma-{\Delta\over2}) = 
	\int\!{dk\over2\pi\hbar}\, 
		e^{{i\over\hbar}k\Sigma} R(k,\Delta)\;.
\end{equation}
From Equation (\ref{propmat}), we find that
\begin{eqnarray}\label{Rpropmat}
R(k,\Delta_f;t) &=& \hbar N(t) \exp\Bigl[-4{\mu k_B T\over\hbar^2}
	\int_0^t\!dt'U(\Delta)\Bigr]\nonumber\\
& & \qquad\qquad\times R\bigl(k,\Delta(0);0\bigr) 
	\Big\vert{\partial \Delta(0)\over\partial k}\Big\vert\;,
\end{eqnarray}
where $\Delta(t')$ is determined by $\Delta_f, k$, and $t$ through
the equation of motion
\begin{equation}\label{EofM2mat}
M\dot{\Delta}(t') - 2\mu U'\bigl(\Delta(t')\bigr) = k\;,
\end{equation}
with the single boundary condition $\Delta(t) = \Delta_f$.  (Whether
one calls this the same equation as (\ref{EofMmat}) seems to be a
matter of semantics.  However one decides the matter, $\Delta(t') =
\Delta(k,\Delta_f,t;t')$ and $\Delta_0(t') =
\Delta_0(\Delta_f,\Delta_i,t;t')$ are closely related:
$\Delta_0(\Delta_f,\Delta_i,t;t') =
\Delta\bigl(K(\Delta_f,\Delta_i,t),\Delta_f,t;t'\bigr)$.)

Evaluating ${\partial\Delta(0)\over\partial k}$ clearly requires
solving Equation (\ref{EofM2mat}).  But we can learn something about
its behaviour by differentiating Equation (\ref{EofM2mat}) with respect
to $k$, keeping $t$ and $\Delta_f$ fixed, to obtain a {\em linear}
equation for ${\partial\Delta(t')\over\partial k}$:
\begin{equation}
M{\partial^2\Delta\over\partial k\partial t'} = 1 + 2\mu U''(\Delta)
	{\partial \Delta\over\partial k}\;.
\end{equation}
The constraint that $\Delta_f$ be held fixed implies the boundary
condition that ${\partial\Delta\over\partial k}\vert_{t'=t} = 0$.  This
equation may then easily be solved, to obtain
\begin{equation}\label{Delta1mat}
{\partial \Delta(0)\over\partial k} = -{1\over M} 
\int_0^t\!dt'\,e^{-{2\mu\over M}\int_0^{t'}\!dt''\, 
	U''(\Delta(t''))}\;.
\end{equation} 

Equation (\ref{Delta1mat}) is easy to evaluate at any fixed point of
Equation (\ref{EofM2mat}).  For example, we know that for $k=0$ there
is fixed point at $\Delta=0$.  We can therefore use (\ref{Delta1mat})
to fix $N(t)$, because the requirement that
$\int\!dx_f\,\rho(x_f,x_f;t) =1$ is equivalent to demanding that
$R(0,0)=1$.  We therefore find that
\begin{equation}\label{normmat}
N(t) = {2\mu U''(0)\over \hbar(1-e^{-{2\mu\over M}U''(0) t})}\;,
\end{equation}
which has the correct dimensions of (length)$^{-2}$.  

The fixed point at the origin of $(k,\Delta)$-space is {\it unstable}.
This is actually a familiar phenomenon, occurring in the
Caldeira-Leggett model\cite{CaldeiraLeggett}:  the fact that a large
range of $\Delta_f$ near the origin are determined by a narrow range of
$\Delta_i$ is precisely what allows the system to ``forget'' its
initial state, and approach equilibrium at late times.  Unstable fixed
points of Equation (\ref{EofM2mat}) are thus easy to associate with
dissipation.  If $U(\Delta)$ were totally parabolic, as in a linear
model, these would be the only fixed points present; but it is easy to
see that if $U$ approaches a constant at large $|\Delta|$, then for
small enough $|k|$ there will also be fixed points that are {\it
stable}.  At these points, the factor $|{\partial\Delta(0)\over\partial
k}|$ in (\ref{Rpropmat}) will {\it grow} exponentially with time.
Careful consideration shows that the case $\hbar^2 U'' > 2 M k_B T U$
in which this exponential growth even overcomes decoherence in Equation
(\ref{Rpropmat}) is actually a violation of our premise that the
thermal frequency $k_BT/\hbar$ is much higher than any other frequency
in the problem.  Nevertheless, the stable fixed points are places where
$R(k,\Delta)$ does not decay as rapidly with time as one might naively
expect.  Their existence is a novel, non-linear phenomenon, whose
interpretation and significance is under investigation.

\subsection{Field models}

We now consider a more realistic case in which a non-linear interaction
between a Brownian particle and its environment causes the decoherence
rate to saturate at large distances.  Here the environment will be a
quantum field in $n$ spatial dimensions.  Because this case is not as
simple as the mattress model, we will only be able to derive certain
properties of the influence functional, but from these we will be able
to draw significant conclusions about the distance-dependence of
decoherence.

Suppose that the interaction Hamiltonian coupling our particle to the
field is of the form
\begin{eqnarray}\label{Hintpf}
H_{int}(t) &=& g\int\!d^ny\,\Phi(\vec{y},t) \tilde{f}\bigl(|\vec{y}-\vec{x}(t)|\bigr)\nonumber\\
&\equiv&\int\!d^ny\,\Phi(\vec{y},t)j(\vec{y},t)\;.
\end{eqnarray}
Here $\vec{x}(t)$ is the position of our Brownian particle (also in $n$
dimensions), and $g$ is a coupling constant.  Note that
$\Phi(\vec{y},t)$ is the quantum field operator in the interaction
picture: the field has a time-independent self-Hamiltonian $H_\Phi$, and 
we have the
interaction picture evolution equation
\begin{equation}\label{IPevol}
i\hbar\dot{\Phi} = [\Phi,H_\Phi]\;.
\end{equation}
Much as in the mattress model above, the particle couples to the field
through a window function $\tilde{f}(|\vec{y}|)$, which has dimensions
of (length)$^{-n}$ and vanishes at large $|\vec{y}|$.  (Our notation
$\tilde{f}$ anticipates the fact that the Fourier transform $f_k$ of
this window function will play essentially the same role as $f_\omega$
in Sections II and III, as long as we use units in which $c=1$ so that
the distinction between spatial and temporal frequency can be made
implicit.)  If $\tilde{f}$ were a delta function, the coupling would be
exactly local; but, to be consistent in neglecting such phenomena as
pair production of more Brownian particles, we will assume that
$\tilde{f}$ has support over some finite UV cut-off length scale.

We again express the evolution of the Brownian particle's reduced
density matrix by Equation (\ref{RDMevol}), with $x\to\vec{x}$ for
$n>1$.  Since any decoherence during this evolution is expressed in the
influence functional, we will focus our attention on
$F[\vec{x},\vec{x}']$.  By assuming that the initial state of the field
$\Phi$ is described by a thermal density matrix 
$\rho_\Phi= Z_\beta^{-1} e^{-\beta H_\Phi}$ uncorrelated with
the initial state of the system, we can write the influence functional
formally as
\begin{eqnarray}\label{IFformal}
F[\vec{x},\vec{x}'] &=& {1\over Z_\beta}
	{\rm Tr}\Bigl\{ 
	{\cal T}\exp\bigl[-{i\over\hbar}\int_0^t\!dt'\,
	H_{int}(t',\vec{x})\Bigr]\nonumber\\
&&\qquad\times \exp\Bigl[-\beta H_\Phi\Bigr]\nonumber\\
&&\qquad\times
 	\bar{\cal T}\exp\Bigl[{i\over\hbar}\int_0^t\!dt'\,
	H_{int}(t',\vec{x}')\Bigr]\Bigr\}\;,
\end{eqnarray}
where $\bar{\cal T}$ denotes reverse time ordering, and the trace is
over the field sector of Hilbert space.

Using the definition of the source field $j(\vec{y})$ from
(\ref{Hintpf}), we can define the {\it influence phase} ${\cal V}[j,j']$, 
such that
\begin{equation}\label{Gamma}
F[\vec{x},\vec{x}'] \equiv \exp i{\cal V}[j,j']\;.
\end{equation}
We have written ${\cal V}[j,j']$ in terms of the sources $j$ instead of
the positions $\vec{x}$ because in this form it is familiar from
quantum field theory as the generating functional for connected
$n$-point functions.  In evaluating $F$ perturbatively in the coupling
$g$, ${\cal V}$ rather than $F$ itself is the most natural object to
compute directly.  It will also be easiest for us to compare ${\cal V}$
with the exponential expressions derived in previous Sections.  In
order to derive illustrative results without undertaking any very
intricate calculations, we will limit ourselves to discussing the
influence phase to second order in $g$.  Assuming that $H_\Phi$ has no
odd-power terms, so that ${\rm Tr}e^{-\beta H_\Phi} \Phi =0$, we find
that this second order term is given by
\begin{eqnarray}\label{Gamma2}
\lefteqn{{\cal V}_2[j,j'] =}\\
&& -{i\over2\hbar^2}
	\int\!d^ny_1d^ny_2\!\int_0^t\!dt_1\!\int_0^{t_1}\!dt_2\,
	\Bigl\{[j(\vec{y}_1,t_1) -j'(\vec{y}_1,t_1)]\nonumber\\
&&\quad\times\Bigl(
	[j(\vec{y}_2,t_2) + j'(\vec{y}_2,t_2)]
	\langle[\Phi(\vec{y}_1,t_1),\Phi(\vec{y}_2,t_2)]\rangle_\beta
	\nonumber\\
& &\qquad - [j(\vec{y}_2,t_2) - j'(\vec{y}_2,t_2)]
	\langle\{\Phi(\vec{y}_1,t_1),\Phi(\vec{y}_2,t_2)\}\rangle_\beta
	\Bigr)\Bigr\}\;,\nonumber
\end{eqnarray}
where $\{A,B\}\equiv AB + BA$, and $\langle A\rangle_\beta \equiv
Z_\beta^{-1}{\rm Tr}\,(e^{-\beta H_\Phi} A)$.

Assuming further that $H_\Phi$ is spatially homogeneous and isotropic,
we can simplify our expressions further by defining the Fourier
transforms
\begin{eqnarray}\label{gdef}
\lefteqn{\langle[\Phi(\vec{y}_1,t_1),
	\Phi(\vec{y}_2, t_2)]\rangle_\beta}\nonumber\\
&& \qquad =i\hbar\int{d^nk\over{(2\pi)^n}}
	e^{i\vec{k}\cdot(\vec{y}_1-\vec{y}_2)}\,G_r(k, t_1-t_2)
	\nonumber\\ 
\lefteqn{\langle\{\Phi(\vec{y}_1,t_1),
	\Phi(\vec{y}_2, t_2)\}\rangle_\beta}\nonumber\\
&&\qquad =\hbar\int{d^nk\over{(2\pi)^n}}
	e^{i\vec{k}\cdot(\vec{y}_1-\vec{y}_2)}\ G_h(k,t_1-t_2)\;,
\end{eqnarray}
where $k\equiv |\vec{k}|$.  Employing also the Fourier transform
$f_k$ of the window function $\tilde{f}(|\vec{y}|)$ from Equation
(\ref{Hintpf}), we can write
\begin{eqnarray}\label{V2x}
\lefteqn{{\cal V}_2[j(\vec{x}),j'(\vec{x'})]=}\nonumber\\
&&\quad -{g^2\over2\hbar}\int{d^nk\over (2\pi)^n}  
f^2_k\int_0^t\!dt_1\!\int_0^{t_1}\!dt_2\,
	\bigl(e^{i\vec k \cdot\vec x(t_1)}
	-e^{i\vec k \cdot\vec x'(t_1)}\bigr)\nonumber\\
&&\qquad\quad\times\Bigl(	
	G_h(k,t_1-t_2)\bigl(e^{-i\vec k \cdot\vec x(t_2)}
		-e^{-i\vec k \cdot\vec x'(t_2)}\bigr)\nonumber\\
&&\qquad\quad
	-iG_r(k, t_1-t_2)\bigl(e^{-i\vec k\cdot \vec x(t_2)}
		+e^{-i\vec k \cdot\vec x'(t_2)}\bigr)
\Bigr)\;. 
\end{eqnarray}

For comparison with our results below, note that the so-called ``dipole approximation'' to (\ref{V2x}), obtained by expanding
to leading order in $\vec{x}-\vec{x}_0$ and $\vec{x}'-\vec{x}_0$
for any constant $\vec{x}_0$, is 
\begin{eqnarray}\label{infldip}
\lefteqn{{\cal V}_{dipole}[j(\vec{x}),j'(\vec{x'})]=}\nonumber\\
&&\qquad -\int_0^t\!dt_1\!\int_0^{t_1}\!dt_2\, 
(\vec x-\vec x')_{t_1}\cdot
	\Bigl((\vec{x}-\vec{x}')_{t_2}\nu(t_1-t_2)\nonumber\\
&&\qquad\qquad\qquad\qquad
	-i(\vec{x}+\vec{x}')_{t_2}\eta(t_1-t_2)\Bigr)\;,
\end{eqnarray}
where the dissipation and noise kernels are given by
\begin{eqnarray}\label{etanu}
\eta(t)&=&{g^2\over 2n\hbar}\int\!{d^nk\over (2\pi)^n}\,k^2
f^2_k G_r(k,t)\nonumber\\
\nu(t)&=&{g^2\over 2n\hbar}\int\!{d^nk\over (2\pi)^n}\,k^2 
f^2_k G_h(k,t)\;.
\end{eqnarray}
Equation (\ref{infldip}) is the familiar form of the influence phase
for a bath of independent harmonic oscillators coupled linearly to a
Brownian particle.

For general $H_\Phi$, it is of course difficult to obtain the complete
propagators $G_r$ and $G_h$.  Formally, however, constraints imposed by
unitarity and causality allow one to write them as
\begin{eqnarray}\label{spectral}
G_r(k, \Delta t)&=&{e^{-\Lambda(k)|\Delta t|}\over 2\omega}
\sin\omega(\Delta t)\nonumber\\
G_h(k, \Delta t)&=&{e^{-\Lambda(k)|\Delta t|}\over 2\omega
(\cosh\beta\omega-\cos\beta\Lambda)}\nonumber\\
&&\quad\times
\bigl(\sinh\beta\omega\cos\omega(t_1-t_2)\nonumber\\
&&\qquad +
\sin\beta\Lambda\sin\omega|\Delta t|\bigr)\;,
\end{eqnarray}
for some $\omega(k,\beta)$ and $\Lambda(k,\beta)$ (which may in
principle be determined by solving Schwinger-Dyson
equations)\cite{LandsmanvanWeert}.  For the purposes of illustration,
we will consider only two simple limiting cases of the dynamics of
$\Phi$: the strongly overdamped case, and the case where $\Phi$ is
free.

The overdamped limit is approached when $\Phi$ is coupled to a large
number of light fields, which are to be traced over as well as (and, by
a purely presentational choice, before) $\Phi$ itself.  The result that
we assume is that $\Lambda(k)$ is, for all important $k$, by far the
highest frequency that is significant in the problem.  Under this
assumption, the exponential decay in the propagators (\ref{spectral})
so dominates their behaviour that they may be approximated by local
distributions, proportional to the delta function or its derivatives.
Thus, the leading contributions to (\ref{spectral}) are found by
setting
\begin{eqnarray}\label{spectral2}
G_r(k,t_1-t_2)&\to&{2\over \Lambda^3(k)}\delta'(t_1-t_2)\nonumber\\
G_h(k,t_1-t_2)&\to&{\delta(t_1-t_2)\over \omega\Lambda(k)}
{\sinh\beta\omega\over (\cosh\beta\omega-\cos\beta\Lambda)}\;.
\end{eqnarray}

Applying (\ref{spectral2}) to (\ref{V2x}), we obtain
\begin{eqnarray}\label{inflloc}
\lefteqn{{\cal V}_2[j(\vec{x}),j'(\vec{x'})]=}\\
&&\quad	-\int_0^t\!dt_1\,\bigl[V_n\bigl(|\vec x-\vec x'|\bigr)
	\nonumber\\
&&\qquad\qquad-iV_d\bigl(|\vec x -\vec x'|\bigr)
	(\dot{\vec x}+\dot{\vec x'})\cdot(\vec{x}-\vec{x}')
	\bigr]\;,\nonumber
\end{eqnarray}
where the functions $V_n(r)$ and $V_d(r)$ are defined to be
\begin{eqnarray}\label{fandg}
V_n(r)&=&-i{g^2\over 2\hbar}\int\!{d^nk\over{(2\pi)^n}} 
{f^2_k\over \Lambda(k)\omega}\nonumber\\
& & \qquad\;\times 
{\sinh\beta\omega\over (\cosh\beta\omega-\cos\beta\Lambda)}
\,(1-\cos{\vec k\cdot\vec r})\nonumber\\
V_d(r)&=&{g^2\over2\hbar r^2}\int\!{d^nk\over{(2\pi)^n}} 
{f^2_k\over \Lambda^3(k)}\vec{k}\cdot\vec{r} e^{i\vec k\cdot\vec r}
\;.
\end{eqnarray}
It is easy to see that, as $r\to 0$, $V_n(r)\propto r^2$ and $V_d(r)$
approaches a constant quadratically.  For $r\to\infty$ on the other
hand, oscillatory terms will wash out in the integrals: $V_n(r)$
approaches a constant, and $V_d(r)\to 0$.  Once again, decoherence
saturates at large distances.

Note that, since Equation (\ref{V2x}) involves a single integral, we
can regard ${\cal V}_2$ as part of an effective action, and derive a
master equation for $\rho(\vec{x},\vec{x}';t)$ by the same method one
uses to obtain the Schr\"odinger equation from the path integral for a
wave function\cite{HPZ2}.  If $H_0(\vec{p}_x,\vec{x})$ is the
self-Hamiltonian for the Brownian particle, the result is
\begin{eqnarray}\label{master}
i\hbar\dot\rho &=& \bigl[H(-i\hbar\vec\nabla_x,\vec x) -
	H(-i\hbar\vec\nabla_{x'},\vec{x}')\bigr]\,\rho \nonumber\\
& &\qquad+ i{\hbar\over M} V_d(|\vec x-\vec x'|)\,
	(\vec{x}-\vec{x}')\!\cdot\!(\vec\nabla_x-\vec\nabla_{x'})\,
	\rho\nonumber\\
& &\qquad -iV_n(|\vec x-\vec x'|)\,\rho\;.
\end{eqnarray}
This is the same form of master equation as that postulated by Gallis in
Reference \cite{Gallis2}.

We now turn to our second simple limit of Equation (\ref{spectral}).
When the field $\Phi$ is free and massless, the propagators have the
following trivial form:
\begin{eqnarray}\label{freeprop}
G_r(k, t_1-t_2)&=&{1\over 2k}\sin k(t_1-t_2)\nonumber\\
G_h(k, t_1-t_2)&=&{1\over 2k}\cos k(t_1-t_2)
\coth\beta\hbar k/2\;.
\end{eqnarray}
In this case, the kernels entering in the influence functional are truly
nonlocal and the behavior is entirely non--Markovian.  Due to 
the interplay between nonlinearity and nonlocality (in time), it is not
possible to obtain a local master equation. 

However, to investigate the behaviour of decoherence as a function of
separation distance, we can evaluate the influence functional for a
pair of simple histories, in which the distance between the two
trajectories remains constant for all times: $\vec x-\vec x'=\vec L$.
In this case the absolute value of the influence functional is
\begin{eqnarray}\label{inflfree}
\lefteqn{|F[x,x']|\equiv \exp\Bigl[{-D_L(t)}\Bigr]}\nonumber\\
&=&\exp\Bigl[-{g^2\over4\hbar}\int_0^t\!dt_1\!\int_0^t\!dt_2 
	\int{d^n\vec k\over (2\pi)^n}
	 {f^2_k\over k}\coth{\hbar\beta k\over2}\nonumber\\
&&\qquad\times \cos k(t_1-t_2)\bigl(1-\cos{\vec k\cdot\vec L}\bigr)
	\Bigr]\;.
\end{eqnarray}

The temporal integration is straightforward, and while for even $n$ the
angular integration produces Bessel functions, for $n=1$ and $n=3$ the
results are tractable integrals over $k$:
\begin{eqnarray}\label{infn1n3}
D_L(t)&\stackrel{\scriptscriptstyle n=1}{=}&
	{2 g^2\over\pi\hbar^2}\ 
	\int_0^\infty {dk\over k^3} f^2_k\nonumber\\
&&\qquad\times  
	\sin^2({k t\over 2})\ coth{\beta\hbar k\over 2}\ 
	(1-\cos k L)\nonumber\\
D_L(t)&\stackrel{\scriptscriptstyle n=3}{=}&
	{g^2\over\pi^2\hbar^2}\ 
	\int_0^\infty {dk\over k} f^2_k\nonumber\\
&&\qquad\times  
	\sin^2({k t\over 2})\ coth{\beta\hbar k\over 2}\ 
	(1-{{\sin k L}\over k L})\;.
\end{eqnarray}
In the convenient case of the Lorentzian window function $f_k^2 =
\Gamma^2/(k^2 +\Gamma^2)$, and in the limits of high temperature or
zero temperature, we can evaluate (\ref{infn1n3}) by using contour
integration (and, in the $n=1$ case, some integration by parts).  At
high temperatures ($k_B T >> \hbar\Gamma$) we obtain
\begin{eqnarray}\label{highT}
\lefteqn{{\hbar^2\Gamma^3\over g^2 k_B T}D_L(t) 
	\stackrel{\scriptscriptstyle T\to\infty}
	{\longrightarrow} (1-e^{-\Gamma L})(1-e^{-\Gamma t})}\\
&&\qquad +\left\{\begin{array}{lr}
		{\Gamma^3 t^2\over2}(L-t/3) 
		- \Gamma t +e^{-\Gamma L}\sinh\Gamma t\;, & t<L\\
		{\Gamma^3 L^2\over2}(t-L/3) 
		- \Gamma L + e^{-\Gamma t}\sinh\Gamma L\;, & t>L
	\end{array}\right.\nonumber
\end{eqnarray}
fpr $n=1$, and
\begin{eqnarray}
\lefteqn{{2\pi\hbar^2\Gamma\over g^2 k_B T}D_L(t) 
	\stackrel{\scriptscriptstyle T\to\infty}
	{\longrightarrow}
	-(1-e^{-\Gamma L})(1-e^{-\Gamma t})}\\
&&\qquad +{g^2 k_B T\over2\pi\hbar^2 \Gamma}
	\left\{\begin{array}{lr}
		\Gamma t - e^{-\Gamma L}\sinh\Gamma t\;,& t<L\\
		\Gamma L - e^{-\Gamma t}\sinh\Gamma L\;,& t>L
	\end{array}\right.\;.\nonumber	
\end{eqnarray}
for $n=3$.

$D_L(t)$ is plotted, for $n=3$ and $T\to\infty$, in Figure 1.  The
shape of the function, being symmetric in $t$ and $L$, vanishing along
the axes, rising with increasing $t+L$, and having a sort of ``ridge''
along the line $t=L$, is qualitatively similar for $n=1$.
\begin{figure}
\begin{center}
	\mbox{\psfig{file=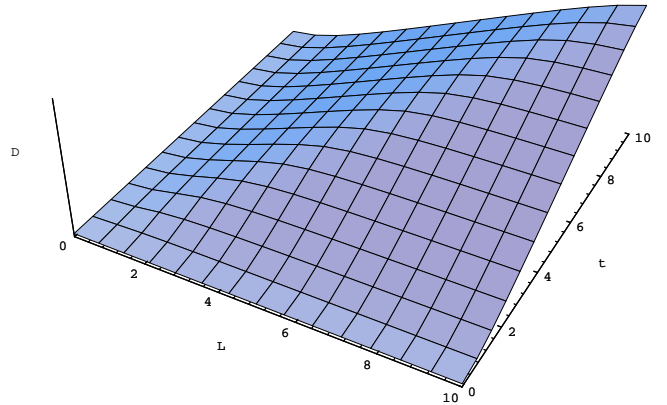,width=8.6cm}}
\end{center}
\caption{The decoherence suppression factor $D_L(t)$, defined as the real
part of the influence phase for two trajectories in which $x$ and $x'$
are constant in time, and differ by $L$.  The environment is a massless
quantum field in $n$ dimensions; the plotted function is for $n=3$ and
high temperature $T\to\infty$.  The $L$ and $t$ axes are in units of
the UV cut-off scale $\Gamma^{-1}$, while the vertical scale is linear but
arbitrary, since it depends on $g^2 k_B T$.} 
\end{figure}

At zero temperature ($\beta\to\infty$) it is convenient to define the 
functions
\begin{eqnarray}\label{kappan}
\kappa_1(z) &=& {1\over2}[e^z {\rm Ei}(-z) + e^{-z} {\rm Ei}(z)]
	-(1 + z^2/2)[C+\ln z]\nonumber\\
\kappa_3(z) &=& C + \ln z 
	- {1\over2}[e^z {\rm Ei}(-z) + e^{-z} {\rm Ei}(z)]\;.
\end{eqnarray}
$C$ is Euler's constant (often called $\gamma$ instead), and ${\rm
Ei}(z)$ is the exponential-integral function\cite{GR}.  In terms of
these functions $\kappa_n$, we have
\begin{eqnarray}\label{zeroT}
D_L(t)\big\vert_{T=0} &=& \kappa_n(\Gamma t) + \kappa_n(\Gamma L)
	 -{1\over2}\kappa_n\bigl(\Gamma (t+L)\bigr)\nonumber\\
&& \qquad 
	-{1\over2}\kappa_n\bigl(\Gamma |t-L|\bigr),\qquad n=1,3\;. 
\end{eqnarray}
For both $n=1$ and $n=3$, the behaviour of $D_L(t)$ is still
qualitatively similar to that shown in Figure 1, even at $T=0$.  The
only noticeable differences are that the ``ridge'' along $t=L$ is
sharper, especially for $n=3$, but that along the top of this ridge the 
function rises somewhat more gradually with increasing $t+L$.
\begin{figure}
\begin{center}
	\mbox{\psfig{file=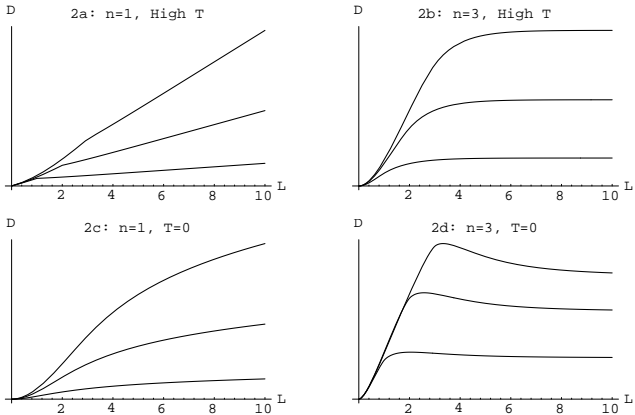,width=8.6cm}}
\end{center}
\caption{
The decoherence suppression factor of Figure 1 plotted versus
$L$ at three successive instants in time: $D_L(m\Gamma^{-1})$, for
$m=1,2,3$ in order from bottom to top, {\it i.e.}, successively higher
curves correspond to later times.  The unit of $L$ is the UV cut-off
$\Gamma^{-1}$.  As in Figure 1, units of $D_L$ are arbitrary, because
each of the four functions plotted has a different prefactor involving
particle-field coupling constant $g^2$ (whose dimensionality depends
on $n$) and/or $T$.  Thus all four vertical scales are linear, but they
are not necessarily commensurate.}
\end{figure}

Figure 2 shows plots of $D_L$ versus $L$ in all four cases, at three
successive instants of time $\Gamma t =1,2,3$.  In each case it is
clear that $D_L$ grows quadratically with $L$ when $L$ is small, but
slows down significantly at large $L$.  For $n=1$, the large $L$
behaviour is linear at high temperatures and logarithmic at zero
temperature; but for $n=3$, $D_L$ actually approaches a constant at
large $L$.  In both cases, a turnover from rapid to slow growth of
$D_L$ can be seen to occur around $L=t$ (although for $n=1$ at high
temperatures this turnover becomes less and less noticeable at later
times).

Even though the functions exhibited in Figures 1 and 2 are not directly
related to the actual behaviour of the Brownian particle (since
trajectories of constant $x$ are unlikely to dominate the path integral
for any $H_0$), they do provide some indication of the dependence of
decoherence on distance, and give a graphic illustration of the
principle which is more firmly established by all the results of this
Section in combination: decoherence does not grow quadratically with
distance in general, but tends to saturate at large distances, in a
manner that will depend in detail on the particular natures of the
environment and its interaction with the system under investigation.

\section{Conclusion}

In general, decoherence is indeed more of a minefield than a checkpoint.  
At low temperatures, and certainly for non-Ohmic environments, decoherence
can be quite complicated even in linear systems.  Noise is coloured,
dissipative terms possess memory, back-reaction can have dramatic effects
even on short time scales, and in general decoherence will be sensitive
to all these features.  With spatial non-linearity, even when noise
is white and dissipation memory-less, decoherence tends to saturate at
long distances, and other novel effects appear.  When non-locality in
time and non-linearity in space are both present, things become still
more complicated, and it is clear that the simple pattern of decoherence
found in Ohmic linear systems at high temperatures is drastically changed.

Since beginning work on this paper, we have become aware of the remarkable
experimental work of Brune {\it et al.}\cite{Brunetal}, in which the
increase of the decoherence rate as the square of the separations scale
is brilliantly confirmed, albeit over a limited range of separations.
Thus, there appear to be sections of the quantum-classical border which
are reasonably orderly.  In this paper, we are paying the current crop
of experiments the highest respect of theorists: we are rushing to keep
ahead of them, by considering still more complicated cases.  And even
so, many of the possibilities we have addressed in this paper seem likely
to be encountered very soon in today's laboratories.

A number of fascinating experiments currently under way are exploring
reaches of quantum physics, such as atom optics, that have been part
of quantum theory since its earliest days, and have been consistently
inferred from observations, but have not hitherto been accessible to
direct empirical investigation.  We certainly expect these experiments
to tell us much about how decoherence occurs in the real world.
But almost all such experiments will be performed at low temperatures,
with non-Ohmic environments and non-linear interactions.  We therefore do
not expect them to confirm the simple formulas that have been obtained
in the first generation of theoretical studies.  Rather, we hope to be
able to use their results to extend our understanding of decoherence
into these more complicated regimes.  Experiments that have recently
been proposed seem to offer yet more scope for investigating hitherto
exotic aspects of decoherence.  In particular, Poyatos, Cirac, and Zoller
have recently shown how one can in principle produce a wide range of
different interaction Hamiltonians between a harmonically trapped ion
and the electromagnetic field\cite{PCZ}.  The future of quantum decoherence
as an experimental study appears to br bright; we will conclude this 
theoretical study with some brief comments on the experimental roles of
the issues we have examined.

The experimental requirement for low temperatures in eliciting
non-classical behaviour is itself evidence supporting the basic
validity of the view that decoherence at high temperatures is what
ensures the effective classicality of the macroscopic world.  At low
temperatures, however, decoherence becomes an interesting phenomenon
in its own right, and not simply a robust mechanism for obtaining
classical behaviour.  In addition to the emergence at low temperatures
of quantum kinematics, one must of course also expect the appearance
of non-trivial quantum dynamics, as lower energy states predominate
and the correspondence principle becomes less powerful.  

Using an internal degree of freedom to enable a classical source
to drive a particle into a Schr\"odinger's Cat state, as in our
Section III, is actually very much what is done in the remarkable
recent experimental construction of a ``Schr\"odinger's kitten'' by
Monroe {\it et al.}\cite{Monroe}.  There are also experiments that use
rather the reverse approach, in which internal degrees of freedom in
the environment are put into superpositions, with the result that a
superposition of two different forces acts on a single system degree
of freedom\cite{Turchette,Haroche}.  It is no co-incidence that both
of these procedures have been suggested for implementing quantum logic,
since the ability to manipulate Cat-like states is the basic requirement
of quantum computing.  Considering decoherence that occurs during
such manipulations, rather than during mere storage of a non-classical
state, is therefore an important task. Our analysis in Section III is
a first step in that direction.  To make it more directly relevant
to the various experiments will require, at the least, extending it
to cases with non-Ohmic environments, in which one might expect to
see non-trivial dependence of decoherence on the time-dependence of
$\alpha(t)$.  For example, one might expect in the case of a supra-Ohmic
environment that if $\alpha(t)$ slowly grows and then shrinks again
to zero, adiabatic dragging would result in decoherence that likewise
rises and then diminishes dramatically.  This possibility of adiabatic
recoherence does not arise to any significant extent in the Ohmic regime.

The current fascinating experiments in atom optics typically
involve local interactions between particles and their
environments\cite{Schmied,Ekstrom}.  One will therefore certainly
expect to see the kinds of saturation effects that we have considered
in Section IV.  Even particles that are free, or confined in simple
enough wells that the dynamics of the particles in isolation is exactly
solvable, are in these cases interacting non-linearly with environmental
degrees of freedom.  This restricted form of non-linearity has not been
extensively studied, and seems capable of providing some interesting
phenomena.  It is also worth noting that, in many experimental set-ups,
one expects environments to be spatially inhomogeneous.  (For example,
in the system of Reference \cite{CTetal} there is an evanescent wave
mirror present only at the bottom of an evacuated cavity.)  This may be
expected to lead to decoherence kernels that are non-trivial functions
not only of off-diagonal variables like the $\Delta$ of our Section IV,
but of mean spatial position as well.

There is clearly a world of experimental possibilities now opening; our
message is that theory must keep up with the times.  We therefore end
with a theorists' proposal for another experiment, in which decoherence
should be adjustable in strength across a wide range.
\vspace{2\baselineskip}
\begin{figure}
\begin{center}
	\mbox{\psfig{file=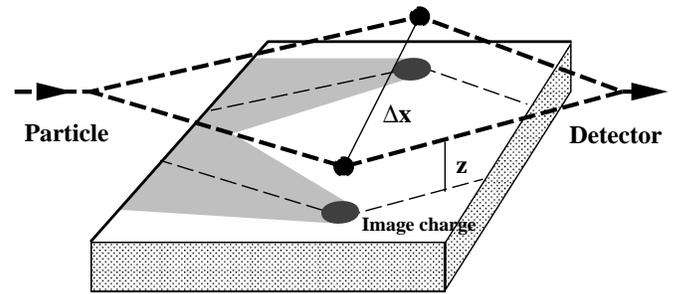,width=8.6cm}}
\end{center}
\caption{
Sketch of proposed system.  The heavy dashed lines indicate two
trajectories of the particle over the conducting plate.  The large shaded
regions represent the disturbance in the electron gas inside the plate.}
\end{figure}

If charged particles are sent through a grating, interference patterns
are the signature of (spatial)
quantum coherence.  This phenomenon is well established, and is
observed consistently as long as the particle beam is isolated from
environmental degrees of freedom.  If an environment is deliberately
introduced, however, in the form of a conducting plate over which the
particles must pass before they are detected, then decoherence may
occur.  A calculation in classical electrodynamics \cite{Boyer} shows
that a charge $Q$ moving at speed $v$ a constant height $z$ above a
plate with resistivity $\rho$ dissipates power a rate
\begin{equation}\label{pwr}
P = {Q^2\rho v^2\over 16\pi z^3}\;.
\end{equation}
This implies Ohmic damping of the particle's motion, with a damping
co-efficient proportional to $\rho z^{-3}$.  Putting a layer of semi-conductor of thickness $b$ on top of the conductor multiplies (\ref{pwr}) by
$2b/3z$\cite{Darling}.  

Since the sensitivity to $z$ is strong, and judicious choice of the
conducting medium permits any $\rho$ from $10^8$ to $10^{-8}$
$\Omega$m, it should be possible to construct an apparatus in which the
effective strength of the system-environment interaction can be varied
so as to span the spectrum between the effectively classical and the
purely quantum regimes.  While the full quantum calculation necessary
to predict the features of decoherence in this system will involve such
complicated quantities as inner products between states of the
conductor's electron gas that have been disturbed by different
trajectories of the particle overhead, the wide variability of the
effective coupling strength should in any case allow one to walk back
and forth across the quantum-classical No Man's Land, exploring it at
leisure.  We are currently considering the theoretical question; we look
forward to being able to compare our results with data from an experiment
along these lines.

\section{Acknowledgements}

J.R.A. would like to thank Salman Habib for valuable discussions.

\vfill

\end{document}